\documentclass{aastex631}
\usepackage[parfill]{parskip}
\usepackage{graphicx}
\usepackage{amssymb}
\usepackage{amsfonts}
\usepackage{amsmath}
\usepackage{epstopdf}
\usepackage{amsthm}
\usepackage{tikz}
\usepackage{pgfplots}
\usepackage[english]{babel}
\usepackage{dsfont}
\usepackage[mathscr]{euscript}
\usepackage{bbm}
\usepackage{bm}
\usepackage{booktabs}
\usepackage[hyphens]{url}
\usepackage{hyperref}

\let\oldcitet=\citet
\let\oldcitep=\citep
\let\oldcitealt=\citealt
\renewcommand{\citet}[1]{\textcolor[rgb]{0,0,1}{\oldcitet{#1}}}
\renewcommand{\citep}[1]{\textcolor[rgb]{0,0,1}{\oldcitep{#1}}}
\renewcommand{\citealt}[1]{\textcolor[rgb]{0,0,1}{\oldcitealt{#1}}}

\received{February 13, 2025}

\graphicspath{{./}{figures/}}
\pgfplotsset{compat=1.18}

\begin{document}
\title{Searching for Low-Mass Exoplanets Amid Stellar Variability with a Fixed Effects Linear Model of Line-by-Line Shape Changes}

\author[0009-0009-4165-9606]{Joseph Salzer}
\affiliation{Department of Statistics, University of Wisconsin-Madison, 1300 University Ave., Madison, WI 53706,
USA; \href{mailto:jsalzer@wisc.edu}{jsalzer@wisc.edu}}

\author[0000-0002-9656-2272]{Jessi Cisewski-Kehe}
\affiliation{Department of Statistics, University of Wisconsin-Madison, 1300 University Ave., Madison, WI 53706,
USA; \href{mailto:jsalzer@wisc.edu}{jsalzer@wisc.edu}}

\author[0000-0001-6545-639X]{Eric B. Ford}
\affiliation{Department of Astronomy \& Astrophysics, 525 Davey Laboratory, 251 Pollock Road, Penn State, University Park, PA, 16802, USA}
\affiliation{Center for Exoplanets and Habitable Worlds, 525 Davey Laboratory, 251 Pollock Road, Penn State, University Park, PA, 16802, USA}
\affiliation{Institute for Computational and Data Sciences, Penn State, University Park, PA, 16802, USA}
\affiliation{Center for Astrostatistics, 525 Davey Laboratory, 251 Pollock Road, Penn State, University Park, PA, 16802, USA}

\author[0000-0002-3852-3590]{Lily L. Zhao}
\thanks{NASA Sagan Fellow} 
\affiliation{Department of Astronomy \& Astrophysics, University of Chicago, 5801 S Ellis Ave, Chicago, IL 60637}
\affil{Center for Computational Astrophysics, Flatiron Institute, 162 Fifth Avenue, New York, NY 10010, USA}

\begin{abstract}
The radial velocity (RV) method, also known as Doppler spectroscopy, is a powerful technique for exoplanet discovery and characterization. 
In recent years, progress has been made thanks to the improvements in the quality of spectra from new extreme precision RV spectrometers. 
However, detecting the RV signals of Earth-like exoplanets remains challenging, as the spectroscopic signatures of low-mass planets can be obscured or confused with intrinsic stellar variability.
Changes in the shapes of spectral lines across time can provide valuable information for disentangling stellar activity from true Doppler shifts caused by low-mass exoplanets. 
In this work, we present a fixed effects linear model to estimate RV signals that controls for changes in line shapes by aggregating information from hundreds of spectral lines. 
Our methodology incorporates a wild-bootstrap approach for modeling uncertainty and cross-validation to control for overfitting. 
We evaluate the model's ability to remove stellar activity using solar observations from the NEID spectrograph, as the sun's true center-of-mass motion is precisely known. 
Including line shape-change covariates reduces the RV root-mean-square errors by approximately 70\% (from 1.919 m s$^{-1}$ to 0.575 m s$^{-1}$) relative to using only the line-by-line Doppler shifts.  
The magnitude of the residuals is significantly less than that from traditional CCF-based RV estimators and comparable to other state-of-the-art methods for mitigating stellar variability.  
\end{abstract}

\section{Introduction}

The radial velocity (RV) method is a method used to discover exoplanets by observing the wobble of a star caused by the gravitational pull of orbiting planets.
This method is not only one of the most prolific for finding exoplanets but also remains essential for accurately constraining their masses.
Using spectrographs, the light from stars is quantified by its stellar spectrum--measurements of flux across different wavelengths.
This spectrum contains dips at specific wavelengths that are referred to as absorption or spectral ``lines," due to the absorption of photons within the stellar atmosphere.
When a star is influenced by an orbiting planet, these absorption lines shift due to the Doppler effect: if $v(t)$ represents the RV at time $t$ and $c$ is the speed of light, then the Doppler-shifted wavelength with respect to a rest wavelength $\lambda$ is given by
\begin{equation}
\label{eq:dopplershift}
\lambda(t) = \lambda \times \left( 1+ \frac{v(t)}{c} \right) \, ,
\end{equation}
\citep{Lovis_2011}.
Observing smooth, periodic RV curves enables astronomers to infer the presence and properties of orbiting planets.

Since the RV-based discovery of 51 Pegasi b \citep{51peg}, astronomers have developed more advanced spectrographs that combine increased resolution, improved instrumental stability, and advanced calibration systems to provide the capacity to identify low-mass exoplanets \citep{Pepe_2013, Jurgenson_2016, Schwab_2016, Seifahrt_2018, Gibson_2018}.
These extreme-precision radial velocity (EPRV) instruments can potentially detect planets akin to Earth.  
A major barrier to Earth-like exoplanet detections is our ability to separate intrinsic stellar variability from true Doppler shifts due to the motion of the host star's center-of-mass (COM) \citep{Crass_2021,Luhn_2023}.  
A planetary Doppler shift causes all lines to shift in the same way without changing spectral line shapes.
In contrast, photospheric activity features---such as spots and faculae---induce line shape changes that can contaminate RV estimates \citep{Hara_2023}.
This adds additional noise on top of expected photon noise and instrument systematics.
These nuisance effects can obscure or mimic planetary signals \citep{Desidera_2004, Queloz_2001, Desort_2007, Robertson_2014}.

A critical step in the analysis of EPRV survey data uses an algorithm to estimate the RV of the COM of each target star (relative to some arbitrary reference) at the time of each observation. 
Given the desire to obtain precise measurements with limited telescope time, information from several hundred to thousands of spectral lines is combined to provide the high signal-to-noise needed to measure small Doppler shifts in the stellar spectrum.  
Traditionally, astronomers first compute the cross-correlation function (CCF) between each stellar spectrum and a fixed ``mask'' that isolates expected line positions;
this effectively averages information from many spectral lines \citep{Pepe_2002}.  
Astronomers then estimate a single RV (and its measurement uncertainty) from the Doppler shift at which the CCF mask best aligns with the observed spectral absorption lines.  
This traditional technique discards information not contained in the mask (which include most of the spectrum) and averages over many lines before estimating RVs.
Another common approach for measuring RVs is based on template fitting \citep{Bouchy_2001, Zechmeister_2018}, which makes use of the precisely known signature of true Doppler shifts.
In this method, estimates of the RV are based on the projection of the residuals onto the derivative of the spectrum (with respect to velocity) and implicitly assumes that any other changes in the spectrum are negligible.  

Recently, astronomers have begun exploring alternative methods for measuring precise RVs that do not begin by averaging thousands of spectral lines or assuming that there are no changes in spectrum shape.  
One such approach is to measure the RV for each spectral line individually \citep{Dumusque_2018,Holzer_2021a}.
In the line-by-line analysis approach, each spectral chunk is analyzed independently, and then line-by-line information is combined.
Each line-by-line RV estimate has a relatively large measurement uncertainty.
One achieves a comparable formal measurement precision to traditional techniques by taking a weighted average of the RVs estimated from hundreds or thousands of lines  \citep{Burrows_2024}.
Some previous line-by-line analyses have focused on selecting which lines to include or reweighting the information from different spectral lines, in an effort to improve the accuracy of the final RV estimate \citep{Cretignier_2020, Moulla_2022}. 
These approaches reduced the contamination from particularly problematic chunks (e.g., chunks with multiple blended lines, telluric contamination, or lines particularly sensitive to magnetic activity).

Other authors have also begun to explore using line-by-line shape information.
For example, \citet{Siegel_2024} also includes line-by-line shape information, using one astrophysically-motivated line shape indicator, and \citet{Holzer_2021a} uses a single line-by-line shape property based on the degree-one Hermite-Gaussian (HG) function which is ultimately combined in a linear model.
In contrast, \citet{Gilbertson_2024} adopts a data-driven approach to analyze entire orders of the \'echelle spectrum without incorporating a priori information about the location of spectral lines.  
Our model adopts an intermediate approach: we identify the locations of lines with a physics-based procedure and then utilize a data-driven technique to estimate line shape changes.
This approach may offer greater interpretability than other line-by-line methods, as we can consider many possible characterizations of line shape, and find those which are most useful for estimating RVs precisely and accurately.

In this study, we develop a fast and flexible approach for estimating precise RVs from spectroscopic time series based on a line-by-line analysis.  
Our model goes beyond previous studies by measuring multiple line shape properties for each spectral line and conditioning on them in the final RV estimate.
As input features, we use summary statistics (e.g., depth and width, coefficients from the projection of each line onto certain basis functions) computed from small wavelength chunks of the extracted spectrum, where each chunk is nearly centered on one of hundreds of known stellar absorption lines.  
The primary outputs are the ``line-by-line RVs'' or the RVs estimated from each selected chunk based on how much the line has shifted relative to some reference spectrum (e.g., the spectrum averaged over all available observations for a given time), while controlling for the considered line shape changes. 

We evaluate our method using NN-explore Exoplanet Investigations with Doppler spectroscopy (NEID) solar observations \citep{Schwab_2016, Gupta_2021, Lin_2022}.
NEID collects over 200 such spectra each clear day and by averaging spectra over each day we obtain high SNR spectra of line-by-line shape analyses.
This approach also averages over solar variability on shorter timescales such as stellar pulsation and granulation, allowing us to focus on longer time-scale effects such as activity features.
We aim to extract and use higher-order shape properties of each spectral line, which generally requires the higher signal-to-noise (SNR) of EPRV spectra.

This manuscript is organized as follows.
The data for our analysis are described in \S\ref{sec:data}, and the proposed modeling regime in \S\ref{sec:method}.
We present the results of several examined models in \S\ref{sec:results}.
Finally, we interpret our results and discuss the implications for future research to improve the extraction of RVs in \S\ref{sec:discuss}.

\section{Data}
\label{sec:data}

We apply our method to data from the NEID spectrograph at the 3.5m WIYN observatory at Kitt Peak National Observatory in Arizona \citep{NEID_optical,NEID_budget}.
The NEID instrument boasts a resolving power of $\sim$110,000 and on-sky, on-star single measurement precision (SMP) of 30 cm s$^{-1}$ for exposures with a SNR of 250 \citep{Gupta_2021}.
While NEID's primary purpose is to survey nearby stars for exoplanets, it also works with the NEID Solar Telescope to take over two hundred spectra of the sun each clear day \citep{Lin_2022}, so as to contribute to research in the development of algorithms for mitigating stellar variability.  
The data analyzed consist of solar measurements collected between 2021 January 1 and 2023 July 7, with some days lacking observations.
Notably, there is a gap in the data between June and November 2022 due to a forest fire. 
For the first few weeks as the spectrograph was brought back online following the forest fire, the RV measurements were poorly calibrated; hence, we have excluded data taken in November of 2022 because of this (a total of 15 days were removed).
Thus, we have 330 days of measurements to include in our model.

The standard NEID data reduction pipeline (DRP v1.2, \url{https://neid.ipac.caltech.edu/docs/NEID-DRP/}) was employed to reduce the stellar spectra for each day of observation, resulting in a time series of wavelength-flux values.
We average spectra collected on the same day to obtain very high SNR and to average over short-term processes such as p-mode oscillations and granulation.  
We identified hundreds of isolated stellar absorption lines using the VALD line list \citep{Piskunov_1995}  avoiding lines identified by a telluric model \citep{TAPAS_2014}.
Our analysis uses wavelength chunks centered on these identified absorption lines.

The NEID instrument records two intensities for some wavelengths, since neighboring \'echelle orders contain overlapping wavelengths.
Consequently, some absorption lines may appear twice in our dataset, originating from different \'echelle orders.
For our analysis, we treat an absorption line that appears on two different orders as separate lines.

We estimate the RV of each line on every date, calculated as a function of the change in the absorption line from a template \citep{Bouchy_2001}, which is computed based on the average flux over all observations for each line. 
The true RV of the COM of the sun, after applying corrections for known Solar System bodies, is expected to be zero.
The primary sources of variability in the measured RVs are therefore intrinsic stellar variability, photon noise, contamination due to telluric absorption, and potentially systematic instrumental variations that were not calibrated out by the NEID DRP.
Thus, we refer to the estimated RV as ``contaminated" by stellar variability and these other sources.  
This paper presents a new approach to computing ``cleaned" RVs that controls for shape changes that could be impacted by stellar variability.

\begin{figure}[htb!]
    \centering
    \includegraphics{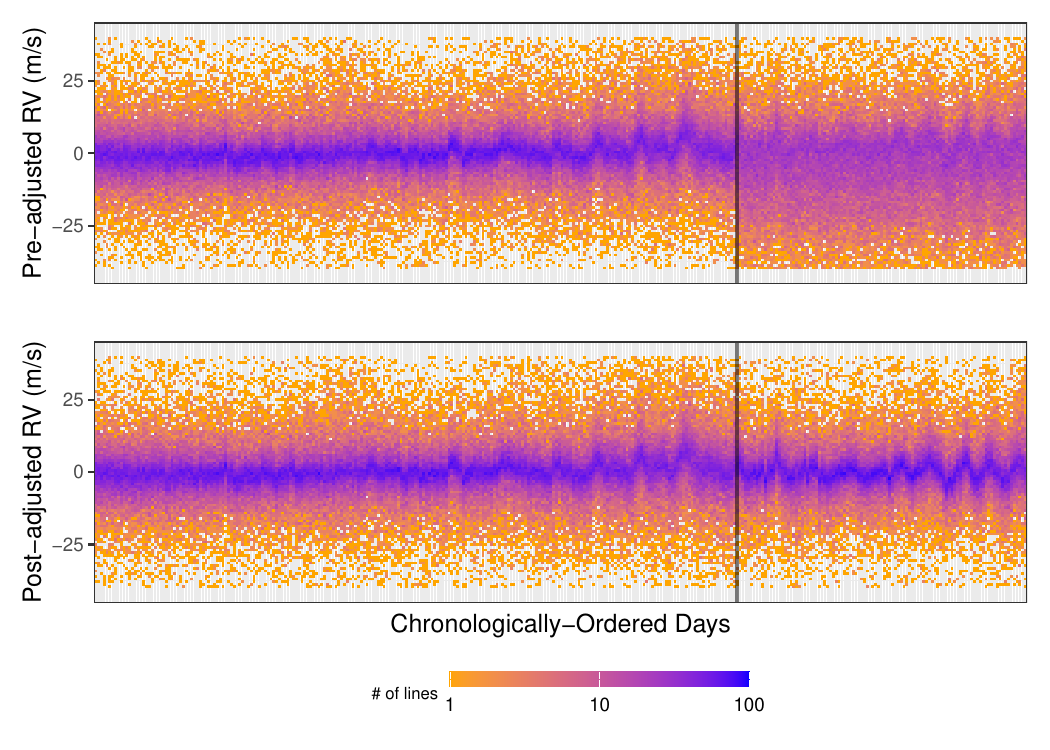}
    \caption{Heatmaps of the NEID Solar RV measurements over time. The x-axis represents the 330 days with measurements (in order, but not scaled by time). The black vertical line is at the day when the detector was shut off due to a forest fire and subsequently brought back online. The top subplot displays the unadjusted distribution of RVs for each line (note that around 4\% of the observations fall outside the limits of this plot). The bottom subplot displays the RVs that have been centered by line and by the temporal group (note that around 2\% of the observations fall outside the limits of this plot). The colors indicate the number of RV measurements from absorption lines within each bin (orange = fewer lines, blue = more lines). Bins are vertical so that there is no overlap between days.}
    \label{fig:heatmap}
\end{figure}

Figure \ref{fig:heatmap} presents a heatmap of the contaminated RV measurements versus time for each line.
This figure highlights a significant distributional difference between the time-points before the fire and after the fire (we refer to these two groups of time-points as ``temporal groups").
Subsequent analysis reveals that this difference is driven mostly by a subset of around 250 of the analyzed absorption lines.
As such, we also display an adjusted version of the RV measurements, where they have been centered by line and temporal group.
In the adjusted plot, the RV distribution appears more similar between the two groups.
Our model is able to inherently adjust for any line-specific and temporal group biases like the one visualized here.

This figure demonstrates the variability in RV measurements, both across lines on the same day and across time.
While many lines have RV measurements that are nearly zero, a significant number of lines have measurements that show substantial variability due to stellar activity and other sources of error.
Additionally, some days exhibit greater variability and bias in the contaminated RV than others.

\begin{figure}[htb!]
    \centering
    \includegraphics{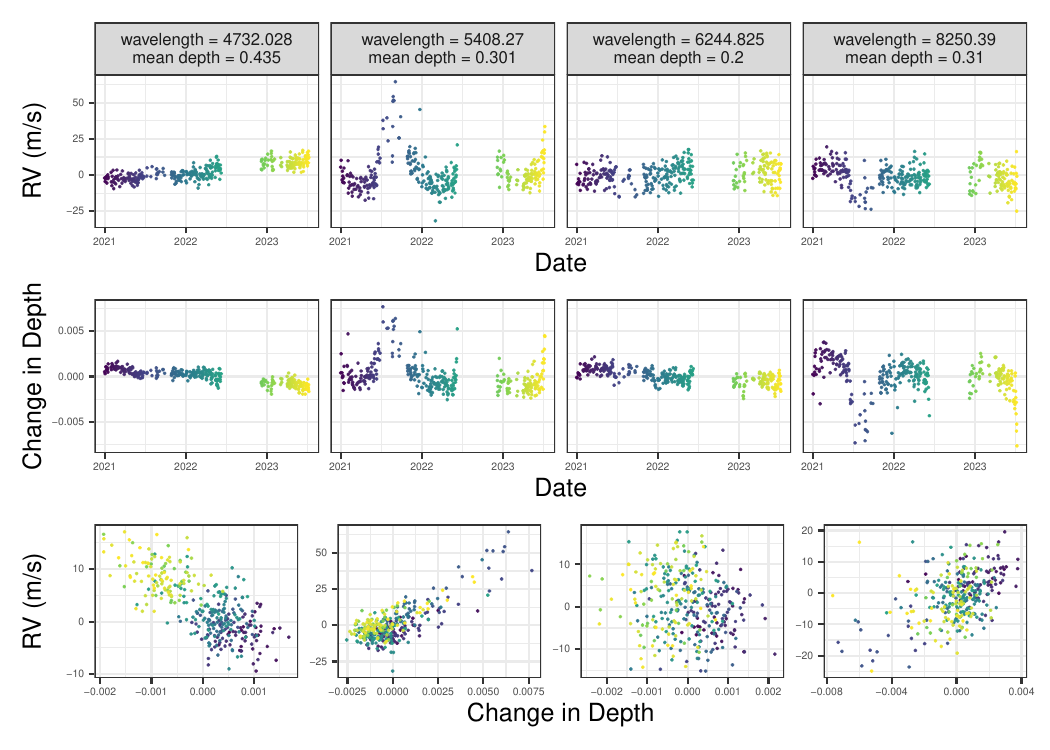}
    \caption{Time series and pairs plots of the contaminated RV and change in depth for four spectral lines. Each column displays results for a single spectral line. The title of each subplot shows the central wavelength (in \AA) of the line followed by its mean depth. The first row displays the contaminated RV, the second displays the change in depth (from the line's mean depth), and the bottom row plots the RV versus the change in depth. Points are colored according to their day of observation.}
    \label{fig:fourLines}
\end{figure}

The time series of contaminated RVs for four individual lines are shown in Figure \ref{fig:fourLines} (top row).
Each line exhibits different patterns, demonstrating that stellar activity (and/or other sources of contamination) affect different lines differently, despite the fact that the true COM RV should be the same for all spectral lines.

Stellar activity can induce variations of line shapes that can mimic a planetary Doppler shift. 
Since stellar activity typically affects line shapes, the evolving shapes of individual spectral lines contain information that can be used to recognize and potentially correct for the presence of stellar activity. 
For example, Figure \ref{fig:fourLines} (middle row) shows changes in the measured line depths over time, and the bottom row shows contaminated RV versus change in line depth; these subplots suggest the two quantities have some degree of correlation.
Importantly, there are differences in the strength and direction of this correlation between the line depth and the measured RV for different lines. 
This demonstrates that there is an opportunity for improving on traditional algorithms for estimating RVs from stellar spectra by leveraging this correlation in a statistical model.

Traditional RV extraction algorithms first ``average'' hundreds of spectral lines (via a cross correlation function) before estimating the RV from the aggregated information.  
Recent work in line-by-line analysis has emphasized filtering more variable lines, so as to improve the quality of RV estimates.
In this paper, we build a model that incorporates the observed correlation of each line's contaminated RV with its estimated depth (and other line shape measurements).
Since a planet affects the RV only through a Doppler shift rather than a shape change, we can utilize the spectral lines' shape changes to obtain information on stellar activity.

Let the indices $l = 1, \dots, L$ represent the spectral lines, and the indices $t = 1, \dots, T$ represent the observation times.
For a small segment of the spectrum centered around a given line $l$ at time-point $t$, we model the normalized flux as a function of wavelength:
\begin{equation}
f_{lt}(\lambda) = [a_{lt}+b_{lt}(\lambda-\lambda_{o,lt})] \left[ 1 - d_{lt} \exp \left( \frac{-(\lambda - \lambda_{c,lt})^2}{2w_{lt}^2} \right) \right] \, ,
\end{equation}
where $a_{lt}$ is the continuum normalization level, $b_{lt}$ is the slope of the continuum, $d_{lt}$ is the fractional line depth, $\lambda_{c,lt}$ is the center wavelength for a Gaussian line profile, $w_{lt}$ is the width parameter for the Gaussian line shape, and $\lambda_{o,lt}$ is a fixed reference wavelength near the center of the region of spectrum being fit\footnote{Here, $\lambda_{o,lt}= \lambda_{o,l}$ is fixed in the assumed stellar rest frame, i.e., accounting for the motion of the observer, but neglecting any planetary perturbations.}.
For each line, this model is fit at every time-point, and the parameters $a_{lt}, b_{lt}, d_{lt},$ and $w_{lt}$ are extracted for use in our statistical model.
This model incorporate knowledge that line shapes are nearly Gaussian, but does not capture deviations from normality.

We also compute the projection of residuals between the observed spectrum and the time-averaged spectrum onto a set of HG basis functions.
This produces a set of coefficients of the HG series expansion for the flux observed around each line at each observation time. 
A change in these scores can be useful for characterizing changes in an absorption line's shape due to activity. 
These have been successfully used to characterize RV scatter in other studies \citep{Holzer_2021b}.
For an idealized Gaussian spectral line, depth changes map onto the zeroth HG function and a pure Doppler shift manifests as adding some of the first HG function (for planetary Doppler shifts that would be small relative to the line width); see \cite{Holzer_2021a} for more details.
For actual spectral lines that are not perfectly Gaussian, these provide useful intuition, but even a pure Doppler shift results in non-zero coefficients for multiple HG functions.
In order to ensure that our line shape measurements are not affected by true Doppler shifts, we analyze only the flux residuals that are orthogonal to a true Doppler shift.

In practice, we first compute a time-averaged spectrum.  We compute the HG coefficients for both the time-averaged spectrum ($\left<H_{lo}\right>_t$) and a version of the time-averaged spectrum ($\left<H^\epsilon_{lo}\right>_t$) that has been Doppler shifted by $v_\epsilon$.  We subtract the two sets of HG coefficients and divide by $v_\epsilon$ to obtain, $d{\bf H}_{l}/dv$, a vector with the derivative of HG coefficients with respect to a pure velocity shift for each line $l$. 
Once we have computed the best-fit HG coefficients for the flux residuals (relative to the time averaged line) at each observation time ($\Delta {\bf H}_{lt}$), then we estimate the Doppler shift by computing the projection of the measured HG coefficients ($\Delta {\bf H}_{lt}$) onto $d{\bf H}_{l}/dv$ and compute the spectral line shape change that is orthogonal to a Doppler shift ($\Delta {\bf H}_{lt,\perp}$) by subtracting off this component:  
$$\Delta {\bf H}_{lt,\perp} =
\Delta {\bf H}_{lt} - \Delta {\bf H}_{lt,\parallel} = 
\Delta {\bf H}_{lt} - \frac{d{\bf H}_{l}}{dv} 
\left(\frac{\Delta {\bf H}_{lt}\cdot \frac{d{\bf H}_{l}}{dv}}{\frac{d{\bf H}_{l}}{dv}\cdot \frac{d{\bf H}_{l}}{dv}}\right).$$
Since the projection onto the HG basis and the projection onto the plane perpendicular to a Doppler shift are linear operations, we can perform these operations on the HG coefficients, rather than on the fluxes themselves, removing the need to interpolate spectra onto a common wavelength grid at each time. 

Before including these line shape measurements (both the orthogonal HG coefficients $\Delta {\bf H}_{lt, \perp}$ and the Gaussian fit parameters $a_{lt}, b_{lt}, d_{lt}, w_{lt}$) in our model, we apply a feature standardization procedure.
For each spectral line and shape property, we subtract its time-averaged shape measurement, and scale these changes so that each shape-change variable has a standard deviation of one across all lines.

Some spectral lines identified using the VALD line list were removed from the dataset. In total, 910 absorption lines were originally identified based on being well-separated from other stellar lines and telluric contamination. 
Lines were excluded if their shape measurements exceeded a threshold used to ensure stable estimates.
Namely, we removed lines if the variable for the slope of the continuum $b$ hit the limit of $\pm.7$, if the width parameter $w$ was exactly the same across all days, or if the Gaussian fit did not converge.
Additionally, lines were removed if they appeared on orders of the detector known to be susceptible to poor wavelength calibration (i.e., physical orders 116, 117, and 118).
Therefore, we analyzed 778 absorption lines from each daily averaged solar spectrum.

The absorption lines we analyzed have a central wavelength coverage from 3896-9892 \AA{} on physical orders 62 to 157.
They include 23 species: Fe 1 (441 total lines), Ni 1 (58), and Si 1 (39) make up the largest groups of line species.
Their mean depths have a interquartile range of $(0.188,0.507)$ and full range of $(0.061,0.854)$, with a median of $0.324$.
The mean SNR by pixel of these lines have an interquartile range of (223.631, 451.320) and a median of 331.708. 

\section{Methodology}
\label{sec:method}

To estimate cleaned RVs, we propose a statistical model.
This model incorporates: a time-varying RV component shared across all spectral lines for each time-point, line-specific RV offsets consistent across all time-points for each lines, and independent variables representing spectral line shape properties (with slopes unique to each line).
The results, presented in \S\ref{sec:results}, evaluate the model's effectiveness in reducing intrinsic stellar variability and instrumental errors.

Consider a linear model, of the form
\begin{align}
\label{eq:contamModel_scalar}
\text{v}_{lt}|x_{lt}  \sim \mathcal{N}(\mu + \alpha_t + \beta_l +  \theta_l x_{lt}, \, \sigma_{lt}^2),
\end{align}
where the response variable, $\text{v}_{lt}$, is the contaminated RV for a given line $l$ with observation time $t$.
We use the template fitting method to estimate these line-by-line contaminated RVs (see \S\ref{sec:data}).
The indices $l = 1, \dots, L$ represent the absorption lines under consideration, and the indices $t = 1, \dots, T$ represent the observation times.

We refer to $\{\alpha_t\}_{t=1}^T$ as ``time fixed effects,'' which are RV offsets associated with each observation time and these are used to derive the cleaned RVs.
The $\{\beta_l\}_{l=1}^L$ are referred to as ``line fixed effects", which are RV offsets for each absorption line.
This offset is necessary because the true source-frame central wavelength of each line is uncertain to $\sim100$ m s$^{-1}$ due to effects such as convective blue shift. 
These offsets are also referred to as ``factors'' since they are discrete variables within our model.
To make the model identifiable, we impose sum-to-zero constraints, $\sum_{t=1}^T \alpha_t = 0$ and $\sum_{l=1}^L \beta_l = 0$.
In \S\ref{subsec:implementation_details} we outline the model in matrix notation, with slight modifications to the time fixed effects to account for differences in RV measurements before and after a time gap due to the forest fire.

Associated with each line at each time we have a covariate, $x_{lt}$, which represents a change in a line's shape--such as its deviation from the mean depth.
While this formulation uses a single covariate, it is generalized to include multiple covariates in \S\ref{subsec:implementation_details}.
The parameters $\theta_l$ are the covariate parameters and allow for each line to have a unique slope relating the change in the covariate to the change in $\text{v}_{lt}$. 
If all lines behaved exactly the same in response to stellar variability and were well separated from all other lines, then a single $\theta$ might be appropriate for all lines.
In practice, each line's response can be affected differently by the slope of the instrumental blaze or stellar continuum around the line, blends with other stellar lines or microtelluric absorption lines, and detector effects.  
Therefore, it may be useful to allow many lines to have their own $\theta_l$. 

The ``noise" of the model is specified by the $\sigma_{lt}^2$ terms, and represents the uncertainty in the RV measurement that results from sources not captured by the rest of the model, such as instrumental errors or unmodeled stellar activity.

A consequence of model \eqref{eq:contamModel_scalar} is that $E( \text{v}_{lt} | x_{lt} = 0 ) = \mu + \alpha_t + \beta_l$ (i.e, the expected RV for line $l$ at observation time $t$ when the shape measurements are at their mean value).
With the sum-to-zero constraints this can be expressed as
\begin{align}
  \label{eq:interpretDecontamRV}
   \frac{1}{L}\sum_{l=1}^L E( \text{v}_{lt} | x_{lt} = 0 ) = \mu + \alpha_t, \\
  \frac{1}{LT}\sum_{t=1}^T \sum_{l=1}^L E( \text{v}_{lt} | x_{lt} = 0 ) = \mu.
\end{align}
Thus $\mu$ (the ``intercept" term) represents the expected RV averaged over all lines and time-points, when the change in the covariates are zero.
Further, the $\alpha_t$ terms represent the difference between the line-wise mean of the expected response at time-point $t$ and the $\mu$.

By controlling for the estimated changes in slope between the contaminated RV and shape of the absorption lines via $\theta_l$, we aim to reduce errors in the contaminated RV estimates that are due to stellar variability.
The $\mu$ and $\beta_l$ terms control for offsets due to the reference frames from which the contaminated RVs are derived.
Simultaneously, we model any remaining time-varying RV signals (e.g., motion due to unknown planets, spurious RV signals due to unmodeled stellar variability) with the $\alpha_t$ terms.
Thus, the $\alpha_t$ terms are used to derive cleaned RV estimates to aid in the detection of possible planetary RV signals. 

Further discussion on the estimation of the cleaned RV and prediction quality of the proposed models is in \S\ref{subsec:est_rvs}.
We outline a cross-validation (CV) procedure in \S\ref{subsec:cv} to quantify how well the model reduces stellar variability and instrumental noise from the contaminated RV.
In \S\ref{subsec:uncertaintyQuantification} we discuss techniques to quantify the uncertainty of the model parameters for our models.

\subsection{Implementation Details}
\label{subsec:implementation_details}

In this paper we explore the capabilities of linear models with time fixed effects to estimate a planetary RV signal by including additional covariates that contain information regarding the absorption lines changes in shape to remove stellar activity.
All proposed models considered in this paper have the following general form:
\begin{align}
\label{eq:contamModel_matrix}
\mathbf{ v } | \mathbf{X_{\text{time}}}, \mathbf{X_{\text{line}}}, \mathbf{X_{\text{covar}}} \sim \mathcal{N}(\bm{\mu} + \mathbf{X_{\text{time}}} \bm{\alpha} + \mathbf{X_{\text{line}}} \bm{\beta} + \mathbf{X_{\text{covar}} }\bm{\theta}, \, \bm{\Sigma} ),
\end{align}
where 
$\mathbf{v}$ (the ``response'') is a vector that contains the line-by-line contaminated RV estimates at each observation time, 
$\bm{\alpha}$ (the ``time fixed effects'') are parameters of the RV offset associated with each observation time,
$\bm{\beta}$ (the ``line fixed effects'') are parameters related to the RV offset associated with each line,
$\bm{\theta}$ (the ``covariate parameters'') are slope parameters associated with the covariates, and
$\mathbf{X_{\text{time}}}, \mathbf{X_{\text{line}}}, \mathbf{X_{\text{covar}}}$ (the ``design matrices'' for the time fixed effects, line fixed effects, and covariates, respectively) are the fixed matrices described below.
For each vector or matrix, we adopt the indices $l = 1, \dots, L$ to represent the absorption lines under consideration, and the indices $t = 1, \dots, T$ to represent the observation times.

An individual response, $\text{v}_{lt}$, is the estimated, contaminated RV for a given line $l$ at observation time $t$; it is contaminated by stellar activity, instrumental errors, and other sources of variability that we seek to remove in order to detect planetary signals. 
The vector form of the response is the $LT \times 1$ vector $\mathbf{ v } = (\text{v}_{11}, \dots, \text{v}_{1T}, \dots, \text{v}_{L1}, \dots, \text{v}_{LT})'$.

The time effects are represented as the $(T-1) \times 1$ vector $\bm{\alpha}$.
This vector contains only $T-1$ parameters to ensure model identifiability.
The cleaned RV time series is estimated by a transformation of the time fixed effects, resulting in estimates for all $T$ time-points (see Equation \ref{eq:cleanRV} below).

Astronomical time-series data frequently contain gaps in their measurements.
In our case, a significant gap occurred when the NEID instrument was shut down for approximately six months due to a forest fire.
To control for possible differences between these two groups of days, we allow each of these groups to have their own offset in our model.
To do this, we must construct the design matrix for the time fixed effects so the model is identifiable.

Suppose our observation times are organized into groups, let $g = 1, \dots, G$ represent the group indices.
Suppose $\Gamma_1, \Gamma_2, \dots, \Gamma_G$ contain sets of sequential time-points $\{1,\dots,T\}$ such that each time-point is contained in exactly one of the sets, $\Gamma_g$.
Finally, let $n_g = |\Gamma_g|$ be the number of time-points in the $g$-th group. such that $\sum_{g=1}^Gn_g = T$.

As noted earlier, sum-to-zero encoding is employed in our model's formulation.
Specifically, we use sum-to-zero encoding for the line fixed effects and a modified version for the time fixed effects to account for differences between data collected before and after the forest fire.
Below, we introduce the general form of the sum-to-zero encoding matrix and its modified version.
Let
\begin{equation}
\label{sum2zero}
\mathbf{S}_n = 
\begin{pmatrix}
1 & 0 & \dots & 0 \\
0 & 1 & \dots & 0 \\
0 & 0 & \dots & 0 \\
\vdots & \vdots & \ddots & \vdots \\
-1 & -1 & \dots & -1 \\
\end{pmatrix},
\end{equation}
be the $n \times (n-1)$ sum-to-zero encoding matrix.

For the time fixed effects, we use a modified version of this encoding to capture offsets for each temporal group as well as offsets of individual time-points relative to their group.
Letting $\mathbf{1}_{n}$ denote a $n \times 1$ vector of ones, we define the encoding for the time fixed effects as
\begin{equation}
\label{eq:modified_contrasts}
\mathbf{S}_{\text{time}} = [ \text{Diag}(\mathbf{1}_{n_1},\dots,\mathbf{1} _{n_G}) \mathbf{S}_G , \, \text{Diag}(\mathbf{S}_{n_1},\dots,\mathbf{S}_{n_G}) ].
\end{equation}
The first term is the $T \times (G - 1)$ matrix that encodes the offset for each group whereas the second term is the $T \times (T - G)$ matrix that encodes the offset of each time-point to that group and serves as our cleaned RV estimates, resulting in $\mathbf{S}_{\text{time}}$ being a $T \times (T-1)$ matrix.

Ultimately, we are interested in the offset of each time-point relative to its group, as this represents the expected RV after accounting for line-shape variations, line-specific offsets, and group-specific offsets.
To capture this, we define the cleaned RVs as the $T \times 1$ vector:
\begin{equation}
\label{eq:cleanRV}
\mathbf{v}^{\text{clean}} = [ \mathbf{0}_{T \times (G-1)}, \, \text{Diag}(\mathbf{S}_{n_1},\dots,\mathbf{S}_{n_G})  ] \bm{\alpha} \, ,
\end{equation}
where $\mathbf{0}_{T \times (G-1)}$ is a $T \times (G-1)$ matrix of zeros.

For our analysis, there are only two temporal groups so $G=2$ with $n_1 = 227$ days before the forest fire and $n_2 = 103$ after the forest fire.
The design matrix for the time effects is given as the $LT \times (T-1)$ matrix $\mathbf{X_{\text{time}}}$; this matrix is $\mathbf{S}_{\text{time}}$ repeated $L$ times row-wise (once for every absorption line).
Specifically, it is $\mathbf{X_{\text{time}}} = \mathbf{1}_{L} \otimes \mathbf{S}_{\text{time}}$, where $\otimes$ is the Kronecker product operator.

The $(L-1) \times 1$ vector $\bm{\beta}$ are RV offsets for each line.
We again use sum-to-zero encoding for these parameters. In particular, the $LT \times (L-1)$ design matrix for the line effects is given as $\mathbf{X_{\text{line}}} = \mathbf{S}_L \otimes \mathbf{1}_{T}$.
The particular form of the design matrices $\mathbf{X_{\text{line}}}$ and $\mathbf{X_{\text{time}}}$ are due to how we have arranged the indices in our response vector $\mathbf{v}$; their structures are similar because each are treated as factors and use similar types of encoding.

If the $\text{v}_{lt}$'s were absolute RV measurements, then the intercept $\bm{\mu} = \mu \mathbf{1}_{LT}$ would be interpreted as the average RV when all changes in covariates are zero, averaged over line and time. 
In practice, $\mu$ is an arbitrary constant, since the $\text{v}_{lt}$'s are differential measurements relative to an arbitrary reference.
This is similar to the offsets we have given to each time group as described in Equation \eqref{eq:modified_contrasts}.
The $\mathbf{v}^{\text{clean}}$ parameters are interpreted as the deviations of the star's RV at each time-point from each group's offset.

For a given model, we estimate $B$ covariates, such as line depth, line width, or the HG basis coefficients.
Throughout this paper, we consider several models that include different sets of these covariates.
We construct the $LT \times LB$ design matrix for these covariates, denoted as $\mathbf{X_{\text{covar}}}$, such that each spectral line has its own shape-change association with the contaminated RV.

The $LB \times 1$ covariate parameters are denoted as
\begin{equation}
\label{eq:covar_params}
\bm{\theta} = (\theta_1^{1}, \theta_2^{1} \dots, \theta_{L}^{1}, \dots, \theta_1^{B}, \theta_2^{B} \dots, \theta_{L}^{B})' \,,
\end{equation}
where the superscript indicates the covariate, and the subscript is the line index.
This structure allows each line to have a unique slope for every covariate.

To illustrate how we construct the covariate design matrix, consider a model that includes only the estimated line depths and widths, so $B=2$.
Denote the depth and width measured for line $l$ at time-point $t$ as $d_{lt}$ and $w_{lt}$, respectively.
Define the $T \times 1$ vectors for line $l$ as $\mathbf{d}_l = (d_{l1}, \dots, d_{lT})'$ and $\mathbf{w}_l = (w_{l1}, \dots, w_{lT})'$.
The $LT\times2L$ design matrix $\mathbf{X_{\text{covar}}}$ is then constructed as:
\begin{align}
\mathbf{X_{\text{covar}}} = [ \text{Diag}(\mathbf{d}_1, \dots, \mathbf{d}_L), \, \text{Diag}(\mathbf{w}_1, \dots, \mathbf{w}_L) ] \,,
\end{align}
where the first term is an $LT \times L$ matrix representing the depths of all lines across time, and the second term is a similar $LT \times L$ matrix for the widths.

The covariance matrix of the conditional distribution of responses is denoted as $\bm{\Sigma}$.
In ordinary least squares (OLS) regression this is assumed to proportional to the identity matrix; in particular, for a scalar noise term $\sigma^2$, $\bm{\Sigma} = \sigma^2 \bm{I}$ (where $\bm{I}$ is the identity matrix).
This reflects the assumptions of homoscedasticity (i.e., each observed RV is drawn from a distribution with the same conditional variance) and zero autocorrelation (i.e., each observed RV is drawn from a distribution that is conditionally independent across time).
Note that the assumption of zero autocorrelation does {\em not} imply temporal independence among the line-by-line RV measurement ($\text{v}_{lt}$'s) if the shape measurement were ignored.
Instead, including shape measurement and time effects can help to mitigate temporal correlations.

The errors are likely heteroscedastic due, for example, to different exposure times or atmospheric conditions across time.
We may also have dependent errors across time due to unmodeled stellar activity or autocorrelated errors even after controlling for shape changes, unaccounted for instrumental errors, etc.
Under these conditions, our OLS estimators are statistically consistent, but not the most statistically efficient -- they do not achieve the smallest possible variance among all linear unbiased estimators \citep{Hayashi_2000}.
In other words, there may be better estimators of the model parameters than the OLS estimators (see \S \ref{subsec:est_rvs}) but as we obtain more data these estimators eventually converge to the true parameters.
However, the OLS estimate for the covariance of the model parameters (see \S \ref{subsec:mle}) is no longer statistically consistent under these general conditions.
We discuss an approach for uncertainty quantification that is more robust to violations of homoscedasticity and zero autocorrelation in \S \ref{subsubsec:bootstrap}.

\subsection{Estimating Cleaned RVs}
\label{subsec:est_rvs}

For notational convenience, let $\mathbf{X} = [\mathbf{1}_{LT}, \mathbf{X_{\text{time}} }, \mathbf{X_{\text{line}} }, \mathbf{X_{\text{covar}} } ]$ be the $LT \times (T + L + LB - 1)$ design matrix for each predictor variable (including the time effects and the line-specific covariates).
Let $\mathbf{x}_k$ be the $(T + L + LB - 1) \times 1$ vector such that $\mathbf{x}_k'$ is the $k$th row of this matrix.
Similarly, let $\bm{\gamma} = [\mu, \bm{\alpha}', \bm{\beta}', \bm{\theta}' ]'$ be the $(T + L + LB - 1) \times 1$ vector of model parameters.
Model \eqref{eq:contamModel_matrix} can then be written as $\mathbf{ v } | \mathbf{X} \sim \mathcal{N}(\mathbf{X} \bm{\gamma}, \, \bm{\Sigma} )$. 

To get cleaned RV estimates we consider the OLS estimators of the model parameters, $\hat{\bm{\gamma}} = (\mathbf{X}'\mathbf{X})^{-1} \mathbf{X}' \mathbf{ v }$.
Recall that we defined the cleaned RV estimates as a linear transformation of the $\bm{\alpha}$ parameters as in Equation \eqref{eq:cleanRV}.
Let $\mathbf{L}$ be the $T \times (T + L + LB - 1)$ linear operator matrix that retrieves all of the cleaned RVs from the model parameters (i.e., this linear operator matrix is given by $\mathbf{L} = [\mathbf{0}_{T \times G}, \text{Diag}(\mathbf{S}_{n_1},\dots,\mathbf{S}_{n_G}), \mathbf{0}_{T \times (L+LB-1)}]$).
The cleaned RV is given as $\mathbf{v}^{\text{clean}} =  \mathbf{L}\bm{\gamma}$, and is estimated as $\hat{\mathbf{v}}^{\text{clean}} =  \mathbf{L}\bm{\hat{\gamma}}$.

One approach for evaluating the cleaned RV estimates, $\hat{\mathbf{v}}^{\text{clean}}$, is to compare them to the true RVs in settings where they are known (e.g., solar data,  simulated planetary signals).
Let $P_t$ be the true planetary RV offset at time $t$; to estimate the root mean square error (RMSE), we use
\begin{equation}
\label{eq:overallRMSE}
\widehat{\text{RMSE}}(\hat{ \text{v} }^{\text{clean}}_t) = \sqrt{ \frac{1}{T} \sum_{t=1}^T  ( \hat{ \text{v} }^{\text{clean}}_t - P_t )^2 } \, .
\end{equation}
This reflects how far off our cleaned RV estimates are from the true COM RV across our dataset.
For our data, we removed the (known) planetary signals and so $P_t = 0$.
If the data had an actual planetary signal, $P_t$ would be the planetary RV offset with respect to the temporal group $t$ of which it belongs.

Models can be compared by their RMSE or other metrics that reflect the ability of the model to predict contaminated RVs.
Let the residuals of the model be given as $\hat{\mathbf{e}} = \mathbf{ v } - \mathbf{X} \hat{\bm{\gamma}}$.
Then, we define the following predictive metrics:
\begin{align}
\text{SSR} &= \hat{\mathbf{e}}' \hat{\mathbf{e}} \\
\text{AIC} &= LT \times \log\left(\frac{\text{SSR}}{LT} \right) + 2 \times (T+L+LB-1) \\
\text{BIC} &= LT \times \log\left(\frac{\text{SSR}}{LT} \right) + (T+L+LB-1) \times \log (LB) \\
\text{RSE} &=  \sqrt{\frac{\text{SSR}}{LT - (T + L + LB - 1)}},
\label{eq:pred_metrics}
\end{align}
known as the sum of squared residuals (SSR), Akaike information criterion (AIC), Bayesian information criterion (BIC), and residual standard error (RSE).
Unfortunately, the RMSE can only be calculated in special cases, such as for simulated data or solar data, where we already know the true planetary RV. 
This motivates our exploration of how well alternative metrics such as AIC or BIC perform for selecting models when applied to night-time observations.

\subsection{Cross Validation}
\label{subsec:cv}

When using solar data, we remove all known planetary signals from the measured solar RVs using the barycentric corrections from \cite{Wright_2020}.
Consequently, throughout our analysis, the planetary RV should remain constant.
This enables us to assess how well we can estimate a planetary signal using cleaned RVs, as discussed in \S\ref{subsec:est_rvs}. 
A general strategy to assess the quality of our model for exoplanet survey targets, where the true RV is unknown, is proposed below.
This CV technique allows for the evaluation of the model's performance on data that was not used for training, and helps to quantify how much of the nuisance RV signal has been removed on unseen data.

In CV, a subset of the data is set aside as a validation set, while the model is trained on the remaining data (i.e., the ``training set'').
The learned parameters are then applied to the validation set to evaluate model performance.
Our dataset has a temporal structure that needs to be accounted for.
Since we are primarily interested in evaluating time-dependent effects, we employ two CV techniques: (1) leave-out-one-day CV, where data from one day is excluded for validation, and (2) leave-out-one-block CV, where blocks of adjacent days spanning various lengths are excluded from validation.

As noted in \S \ref{sec:data}, the data are standardized prior to model fitting.
When applying the model to the validation set, we avoid using any information from that set to standardize the training data.
This includes the covariates’ means (for each line) and standard deviations (across all lines).
Standardization procedures are thus applied solely to the training data and carried over to the validation set.

\subsubsection{Leave-Out-One-Day Cross Validation}
\label{subsubsec:loodcv}
For our leave-out-one-day CV (LOODCV) procedure, let $\mathbf{X}^{(-t)}$ be the design matrix from \S\ref{subsec:est_rvs}, excluding data from time-point $t$, which serves as the validation set.
The model parameters $\bm{\gamma}^{(-t)} = [ \mu^{(-t)}, \bm{\alpha}^{(-t)'}, \bm{\beta}^{(-t)'}, \bm{\theta}^{(-t)'} ]'$ are estimated as $\hat{\bm{\gamma}}^{(-t)}$ via OLS on the training data.
These parameters are then used to make predictions on the validation set.

Define $\mathbf{X_{\text{line}} }^{(t)}$ and $\mathbf{X_{\text{covar}} }^{(t)}$ as the design matrices associated with line effects and covariates, respectively, with dimensions $L \times (L-1)$ and $L \times LB$.
Let $\mathbf{v}^{(t)} = (\text{v}_{1t}, \text{v}_{2t}, \dots, \text{v}_{Lt})'$ be the contaminated RV responses for the left-out time-point $t$.

Since time effects are treated as factors, no parameter in the training design matrix $\mathbf{X}^{(-t)}$ corresponds to the omitted time-point (i.e., $\bm{\alpha}^{(-t)}$ is a $(T-2)\times 1$ vector). 
If the data contain temporal groups and we use the modified contrasts of Equation \eqref{eq:modified_contrasts} to construct the training design matrix, we account for offsets for different groups but not the omitted time-point.
For the left-out time-point $t$, let $\mathbf{X_{\text{time}} }^{(t)}$ represent the $L \times (T-2)$ design matrix associated with the group-wise offset. Details on its specification are in \S \ref{subsubsec:loobcv}.

Consider the difference between the observed contaminated RVs and the predicted contaminated RVs for time-point $t$.
For the validation set, the contaminated RV is predicted and subtracted from the observed contaminated RV as
\begin{equation}
\label{eq:decontamRV_cv}
\tilde{ \mathbf{v} }^{(t)} = \mathbf{v}^{(t)} - (\hat{\mu}^{(-t)} \mathbf{1}_{L} + \mathbf{X_{\text{time}} }^{(t)} \hat{\bm{\alpha}}^{(-t)} + \mathbf{X_{\text{line}} }^{(t)} \hat{\bm{\beta}}^{(-t)} + \mathbf{X_{\text{covar}} }^{(t)}\hat{\bm{\theta}}^{(-t)} ),
\end{equation}
which is a $L \times 1$ vector.
We call $\tilde{ \mathbf{v} }^{(t)}$ the ``line-by-line cleaned RVs";
it represents the expected RVs (on the validation set) after controlling for the covariates and line fixed effects that are estimated using the training set.

This process is repeated for every time-point in our dataset, $t = 1, \dots, T$, resulting in a set of line-by-line cleaned RVs across time.
We can then use these $L$ line-by-line cleaned RVs to derive a single, overall cleaned RV at a given time-point.
A simple choice for the overall cleaned RV is the mean of the line-by-line cleaned RVs, such that our overall cleaned RV at time-point $t$ is given as $\frac{1}{L} \sum_{l=1}^L \tilde{\text{v}}_l^{(t)} $.
We can assess model performance by using Equation \eqref{eq:overallRMSE} to compare our cleaned RVs to the true planetary signal when it is known.

To quantify model performance when the true planetary signal is unknown, we compute the standard error between the line-by-line cleaned RVs and its overall cleaned RV:
\begin{equation}
\label{eq:standerror_cv}
\widetilde{\text{se}}^{(t)} = \sqrt{ \frac{1}{L-1} \sum_{l=1}^L \left( \tilde{\text{v}}_{l}^{(t)} - \frac{1}{L} \sum_{l=1}^L \tilde{\text{v}}_l^{(t)} \right)^2 } \, .
\end{equation}
Here, $\widetilde{\text{se}}^{(t)}$ represents the spread between the overall cleaned RV and the line-by-line cleaned RVs at time-point $t$.
We compare the standard error across different models in order to determine the extent to which the stellar activity has been mitigated from the RVs via the change in shape parameters.

\subsubsection{Leave-Out-One-Block CV}
\label{subsubsec:loobcv}
While the LOODCV procedure offers robust measurements of predictive performance and standard error, autocorrelation in consecutive time-points could still impact the analysis.
To address this, we extend the CV approach by using larger blocks of days for validation, ensuring the model is trained on data that is temporally distant from the validation set.

A similar procedure to LOODCV can be applied.
Instead of removing a single day, we remove overlapping sequences of days.
For instance, we could use a block size of 14 days for our validation set. For each day in the dataset, we remove all other days within 7 days of it, train the model on the remaining days, and evaluate the model only on the central day.
It is important to note that the number of time-points in each validation set depends on the specific days for which we have observations, but we only test the model on the central day.
For example, if observations were made on days 1, 3, 8, and 9, the first validation set for a block size of 14 would include data from days 1 and 3 (with the model being tested on day 1).
The second validation set would include days 1, 3, 8, and 9 (testing on day 3).
The third validation set would consist of days 3, 8, and 9 (testing on day 8).
This process continues for each observed day in the dataset.

The solar rotation rate varies with latitude.
For latitudes that commonly have active regions, the rotation period is approximately one month, so we could also consider leaving out 28 days for the validation set, a method referred to as ``leave-out-one-month" CV (LOOMCV).
For other stars, it might be useful to leave out blocks corresponding to the approximate rotation period.
In each case, the size of the blocks corresponds similarly to the $p$ parameter in leave-$p$-out CV.
As in \S \ref{subsubsec:loodcv} we can compare the contaminated RV with the line-by-line cleaned RVs of the validation set in these multi-day CV procedures.
The RMSE for each central day can again be calculated in a similar manner, and each of these CV regimes has its own RMSE based on the cleaned RV compared to the planetary RV.

Let $(t_a,\dots,t_b)$ be one of the validation sets for an arbitrary leave-one-block-out CV, with size $T_{\text{test}}$.
Following \S \ref{subsubsec:loodcv} we define $\mathbf{X}^{-(t_a,\dots,t_b)}$ as the design matrix from \S\ref{subsec:est_rvs}, excluding data from time-points $(t_a,\dots,t_b)$.
All steps from the LOODCV procedure extend naturally to leave-one-block-out CV.
However, for clarity, we explicitly specify the $LT_{\text{test}} \times (T-T_{\text{test}}-1)$ design matrix for the time effects, denoted as $\mathbf{X_{\text{time}} }^{(t_a,\dots,t_b)}$.
If the data is partitioned into $G$ temporal groups, let $\mathbf{S}_{\text{test}}$ be the $T_{\text{test}} \times (G-1)$ matrix given by extracting the $t_a$ through $t_b$ rows of $\text{Diag}(\mathbf{1}_{n_1},\dots,\mathbf{1} _{n_G}) \mathbf{S}_G$; this matrix represents the temporal group memberships of the validation set time-points.
Then, the design matrix for the time effects will be given by $\mathbf{X_{\text{time}} }^{(t_a,\dots,t_b)} = \mathbf{1}_L \otimes \left[ \mathbf{S}_{\text{test}} \, , \mathbf{0}_{T_{\text{test}} \times (T- T_{\text{test}} -G)} \right]$.

\subsection{Uncertainty Quantification}
\label{subsec:uncertaintyQuantification}

We estimate uncertainties in our model parameters using two different methods. First, when using OLS to calculate point estimates of the parameters, we can estimate the covariance matrix of the model parameters.
In doing so, we compute standard errors for various parameter estimates of interest, including the cleaned RVs. 
However, these require a correct specification of the model covariance matrix $\bm{\Sigma}$ in Equation \eqref{eq:contamModel_matrix}.
Our data may violate homoscedasticity due to both photon noise (which can be approximated easily) and unmodeled sources of error (which are difficult to approximate).
Indeed, we may also have correlated RV measurements across time-points for individual lines.
Either would violate the assumption that $\bm{\Sigma}$ is proportional to the identity matrix.
Therefore, we relax this assumption and carryout a different approach to uncertainty quantification using a bootstrapping technique.

\subsubsection{Maximum Likelihood Estimation}
\label{subsec:mle}

Under a general $\bm{\Sigma}$, the covariance of the estimated parameters--given by $\text{Cov}(\hat{\bm{\gamma}}) = (\mathbf{X}' \mathbf{X})^{-1} \mathbf{X}' \mathbf{\Sigma} \mathbf{X} (\mathbf{X}' \mathbf{X})^{-1} $--is difficult to estimate because it requires a precise estimate of the model covariance $\bm{\Sigma}$.
Putting restrictions on $\bm{\Sigma}$, such as that its diagonal with a scalar noise term, allows for a simplified estimate of the covariance of the estimated parameters.
The noise term, $\sigma^2$, is estimated via $\hat{\sigma}^2 = \text{SSR}/LT$, i.e. the sum of squared residuals divided by the total number of observations.
The covariance of the model parameters is $\text{Cov}(\hat{\bm{\gamma}}) = \hat{\sigma}^2 (\mathbf{X}' \mathbf{X})^{-1}$.
This is utilized to find the variance of the estimators $\hat{\mathbf{v}}^{\text{clean}}$ as $\text{Cov}( \mathbf{L}\hat{\bm{\gamma}}) = \hat{\sigma}^2 \mathbf{L} (\mathbf{X}' \mathbf{X})^{-1} \mathbf{L}'$. 

One consequence of the MLE approach is that each $\hat{\text{v}}^{\text{clean}}_t$ have similar variances.
This is due to how we treat the time fixed effects as factors.
Since the errors are assumed to have constant variance across time and $\mathbf{X}$ is structured similarly for each day, the variance of the fixed effects $\hat{\alpha}_t$ are very close (they are not exactly the same due to the inclusion of the continuous covariates in our model).
This likely does not reflect the true data generating process, as RV measurements likely have different uncertainties on different days. 

Another consequence is that it is difficult to derive the distributional properties of other statistics that might be of interest, namely the RMSE that is estimated in Equation \eqref{eq:overallRMSE}.
We are interested in deriving a confidence interval for this statistic as opposed to a point-wise estimate derived from a single instance of our data; the MLE approach is not conducive to this type of estimate.

The consistency of the MLE approach relies on key assumptions of the variance in model \eqref{eq:contamModel_matrix}.
In particular, that  $\bm{\Sigma} = \sigma^2 I$.
If this is not satisfied, the standard error of the cleaned RVs may not be correct.
Thus, we next consider alternative approaches to derive standard error estimates that relax these assumptions.

\subsubsection{Wild Bootstrap}
\label{subsubsec:bootstrap}

The ``wild bootstrap" may be used to derive standard errors for parameters of interest \citep{Liu_1988}.
This method is particularly useful in regression models with heteroscedastic errors and can be extended to account for time-correlated observations.
The wild bootstrap involves generating an ensemble of simulated datasets by resampling the residuals of the estimated model to create new samples of the responses.
The model is then re-fit to each resampled dataset, and estimates of the model parameters (or any test statistic of interest) are extracted for each resample to build its empirical sampling distribution.

To perform the wild bootstrap, we first fit the model on the original dataset to obtain the model parameters $\hat{\bm{\gamma}}$ and residuals $\hat{\mathbf{e}} = ( \hat{e}_{1},\dots, \hat{e}_{LT})$.
Choose $R$ as the number of bootstrap samples to conduct. Following \citet{DAVIDSON_2008}, we consider a transformation of the residuals using their leverage.
The leverage $h_k$ of the $k \in \{1, \dots, LT \}$ observation is given by $h_k = \mathbf{x}_k' (\mathbf{X}' \mathbf{X})^{-1} \mathbf{x}_k$.
The leverage is a measure of how influential a single observation is to the estimated model, and is used to help uncover outliers.
The residuals are transformed as $\hat{e}_k/(1-h_k)$.

For each $r = 1, \dots, R$, bootstrapped residuals $\mathbf{e}_r^\star$ are generated by multiplying the transformed residuals by an independent random variable $u_{k}$ drawn from a Rademacher distribution (i.e, $u_{k} = \pm 1$ with equal probability).
Specifically, $e^\star_{k,r} = u_k \cdot \hat{e}_k/(1-h_k)$.
We construct the bootstrapped response as $\text{v}^\star_{k,r} = \mathbf{x}_k' \hat{\bm{\gamma}} + e^\star_{k,r}$.
This is simply the original prediction plus a bootstrapped residual, used to generate new responses $\mathbf{v}_r^\star =  (\text{v}_{1,r}^\star, \dots \text{v}_{LT,r}^\star)'$.

We re-estimate the model parameters using the original design matrix $\mathbf{X}$ and bootstrapped responses $\mathbf{v}^\star$. 
Let the $r$th bootstrap estimate be given as $\bm{\gamma}^\star_r = (\mathbf{X}'\mathbf{X})^{-1}\mathbf{X}'\mathbf{v}_r^\star$.
Using all $R$ bootstrap estimates, standard errors can be derived for the cleaned RVs using the standard deviation of the bootstrapped estimates.
Bootstrapped estimates for other statistics of interest, like the RMSE, can be found in a similar manner. 

This procedure assumes conditional independence of the contaminated RV estimates, meaning that $\bm{\Sigma}$ from model $\eqref{eq:contamModel_matrix}$ is diagonal.
This is a significant assumption, ensuring that the resampling procedure accurately reflects the original data generation process.
While using time and line fixed effects controls for time-varying and line-specific RV signals, it may not entirely eliminate conditional dependence.

To capture dependence structures within blocks of data, the wild bootstrap can be modified by using blocking.
Blocking involves dividing the data into blocks where observations are not independent, such as an autocorrelated time series of an absorption line's RV measurements.
Instead of generating bootstrapped residuals for every observation, the observations are grouped by line as $k_l = \{1+(l-1)T, \dots, T+(l-1)T\}$ for $l \in \{1,\dots, L\}$.
Following the same procedure as above, a Rademacher random variable $u_{k_l}$ is generated and used for every bootstrapped residual of the same line.
This ensures that the bootstrapped responses are correlated within a line's observations but are independent between lines.
Ultimately, any blocking structure can be considered where observations are expected to be dependent within a block but independent between blocks, leading to more accurate inferences.

\section{Results}
\label{sec:results}

We organize our results as follows: In \S \ref{subsec:full_results}, we examine the performance of several models using the entire dataset as input.
In \S \ref{subsec:cv_results}, we present cross-validation results for a subset of these models.
Finally, in \S \ref{subsec:boot_results}, we provide the cleaned RV uncertainty quantification results for the best-performing model.

\subsection{Full-data Model Performance} \label{subsec:full_results}
We evaluate several models of the form given in model \eqref{eq:contamModel_matrix}, varying the number of shape change measurements included.
For context, we compare these models to a ``Baseline Model," which excludes all shape change variables and only includes the time and line fixed effects.
This Baseline Model provides a simple estimate of the RV without accounting for line-specific shape changes; its cleaned RV is roughly equivalent to the empirical mean of the contaminated RV across all lines at each time-point.

The shape change variables come from two fits to the spectral lines: the Gaussian fit parameters (continuum normalization level $a$, slope of the continuum $b$, fractional line depth $d$, and width $w$) and the HG coefficients from 0 to 10 (excluding 1).
These sets of variables are labeled ``all" or ``none" depending on which of the variables are included.
For completeness, we label the model which includes all of the Gaussian fit variables and HG coefficients as the ``Full Model."

Another comparison model is the ``Common Slopes Model." Recall that our initial representation of the model in Equation \eqref{eq:contamModel_scalar} specified a unique slope for each line (and each shape change variable).
Instead of this, we could force each line to have a single, common slope for every covariate.
The Common Slopes Model presented in Table \ref{tab:model_comparison} uses all of the Gaussian fit and HG covariates.

To encourage interpretability and reduce the risk of overfitting to noise, we apply L1 regularization--commonly referred to as the least absolute shrinkage and selecton operator (LASSO)--to the Full Model \citep{Tibshirani_1996}. 
Specifically, we use all of our shape change variables and include an L1 penalty to all of the covariate parameters $\bm{\theta}$ (see Equation \eqref{eq:covar_params}) for the least-squares objective.
The regularization parameter was chosen by minimizing the AIC.
For this optimal parameter, we determine which of the covariate parameters are exactly zero and re-fit the (non-regularized) model using only those covariates which are not zero.
We label this model as the ``Full Model w/ LASSO."

To compare our models, we consider their predictive performance via the AIC, BIC, and RSE. Since we know the true RV, we also include the estimated RMSE as described in Equation $\eqref{eq:overallRMSE}$.
Table \ref{tab:model_comparison} summarizes the results for all models based on the full dataset.

\begin{table}[htb!]
\centering
\caption{Model Comparison}
\begin{tabular}{lccccc}
\toprule
\textbf{Model} & \textbf{Parameters} & \textbf{AIC} & \textbf{BIC} & \textbf{RSE} & \textbf{$\text{RMSE}$} \\
\midrule
Baseline Model & 1,107 & $17.061 \times 10^5$ & $17.177 \times 10^5$ & 27.674 & 1.919 \\
Common Slopes Model & 1,121 & $16.335 \times 10^5$ & $16.452 \times 10^5$ & 24.024 & 2.145 \\
Gauss=all\_HG=none & 4,219 & $14.122 \times 10^5$ & $14.563 \times 10^5$ & 15.520 & 1.818 \\
Gauss=none\_HG=all & 8,887 & $12.911 \times 10^5$ & $13.841 \times 10^5$ & 12.152 & 1.745 \\
Full Model & 11,999 & $9.938 \times 10^5$ & $11.192 \times 10^5$ & 6.770 & 0.575 \\
Full Model w/ LASSO & 11,581 & $9.934 \times 10^5$ & $11.145 \times 10^5$ & 6.771 & 0.581 \\
\bottomrule
\end{tabular}
\label{tab:model_comparison}
\end{table}

The Baseline Model, which excludes shape change variables, performs the worst in all predictive metrics.
It does not account for stellar activity, resulting in high residuals and RMSE.
The Common Slopes Model improves the predictive performance but produces worse RMSE estimates due to its inability to model line-specific activity.
Models including either Gaussian variables or HG coefficients alone outperform the Baseline Model.
However, combining these covariates in the Full Model leads to superior results, with an RMSE reduction of $\sim 70\%$ compared to the Baseline Model (from 1.919 m s$^{-1}$ to 0.575 m s$^{-1}$).

The Full Model w/ LASSO removed 418 covariates ($\sim 4\%$ of the covariate parameters).
Compared to the Full Model, the LASSO model had smaller AIC and BIC metrics but had slightly higher RSE and RMSE metrics.
We also fit a LASSO Model where the regularization parameter was chosen by minimizing the BIC (instead of the AIC).
This procedure produced a model that removed 1,099 covariates ($\sim 10\%$ of the covariate parameters), had a smaller BIC (of $11.091 \times 10^5$) but had larger values for AIC and RSE; further, it had a higher RMSE of 0.620.

\begin{figure}[htb!]
    \centering
    \includegraphics{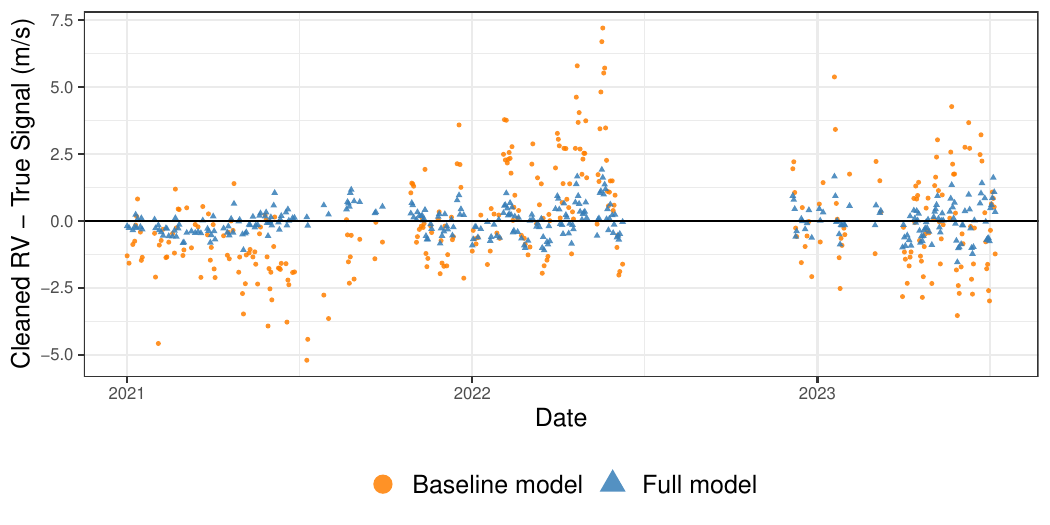}
    \caption{Time series of the cleaned RV minus true planetary signal for the Baseline Model (orange circles) and Full Model (blue triangles). For this dataset, the true planetary signal is $0$ m s$^{-1}$.}
    \label{fig:cleanRV}
\end{figure}

Figure \ref{fig:cleanRV} plots the cleaned RV minus the true planetary signal for the Baseline and Full Models.
The Full Model’s estimates closely align with the true signal, in stark contrast to the Baseline Model.
Note that the true planetary signal here is $0$ m s$^{-1}$, since we use solar data with the known planetary signals removed; however, if we inject any planetary signal into our data, this figure and all metrics in Table \ref{tab:model_comparison} remain the same.

For the examined models, smaller AIC, BIC, and RSE statistics--which assess the model's ability to predict the contaminated RVs--are generally associated with lower RMSE values, which quantifies the model's ability to estimate the cleaned RVs.
This is a potentially useful finding, in that a better predictive model tends to correspond to a model that more accurately estimates the COM motion of the sun.
Though our data do not include any planetary signal so the sun's COM motion is 0 m s$^{-1}$, this still may be useful for other target stars with unknown RVs; however, highly predictive (in terms of lower AIC, BIC, or RSE) models might also mask a planetary signal and thus this relationship may not necessarily hold for all models or data.
We recommend practitioners only include covariates in a model that are invariant to Doppler shifts to avoid masking the planetary signal.

\subsection{CV Results}
\label{subsec:cv_results}

To evaluate the predictive performance of our models, we conducted CV procedures as described in \S\ref{subsec:cv}.
This approach provides insights into the robustness of the models when applied to time-points excluded from the training set.
We employed three CV schemes with varying temporal windows: leave-out-one-day CV (LOODCV, one-day window), leave-out-one-week CV (LOOWCV, 14-day window), and leave-out-one-month CV (LOOMCV, 56-day window).
Since measurements are not available for every day, the validation sets for LOOWCV and LOOMCV do not always span identical numbers of days.
Instead, the window sizes indicate the temporal distance between the test day and training data.

The RMSE for these procedures, summarized in Table \ref{tab:rmse_cv}, compares the performance of the Baseline Model, Full Model, and Full Model with LASSO.
Specifically, the cleaned RV for each test day is calculated as the mean of the line-by-line cleaned RVs (see Equation \eqref{eq:decontamRV_cv}), and this value is used to compute the RMSE (see Equation \eqref{eq:overallRMSE}).

\begin{table}[htb!]
\centering
\caption{RMSE by CV Procedure}
\begin{tabular}{lccccc}
\toprule
\textbf{Model} & \textbf{LOODCV} & \textbf{LOOWCV} & \textbf{LOOMCV} \\
\midrule
Baseline Model & 1.930 & 1.967 & 2.026 \\
Full Model & 0.580 & 0.594 & 0.612 \\
Full Model w/ LASSO & 0.585 & 0.600 & 0.618 \\
\bottomrule
\end{tabular}
\label{tab:rmse_cv}
\end{table}

The Baseline Model consistently produces higher RMSE values across all CV procedures, reflecting poor predictive performance relative to the other models.
In contrast, the Full Model achieves lower RMSE values across all CV procedures, approximately $70\%$ smaller compared to the baseline.
This demonstrates its ability to effectively incorporate line shape information for improved cleaned RV estimates.
While there is a modest increase in RMSE as the window size grows,
they are relatively stable across time gaps.
Similarly, the Full Model with LASSO yields RMSE values close to the Full Model, with slightly higher RMSEs for all procedures.

\begin{figure}[htb!]
    \centering
    \includegraphics{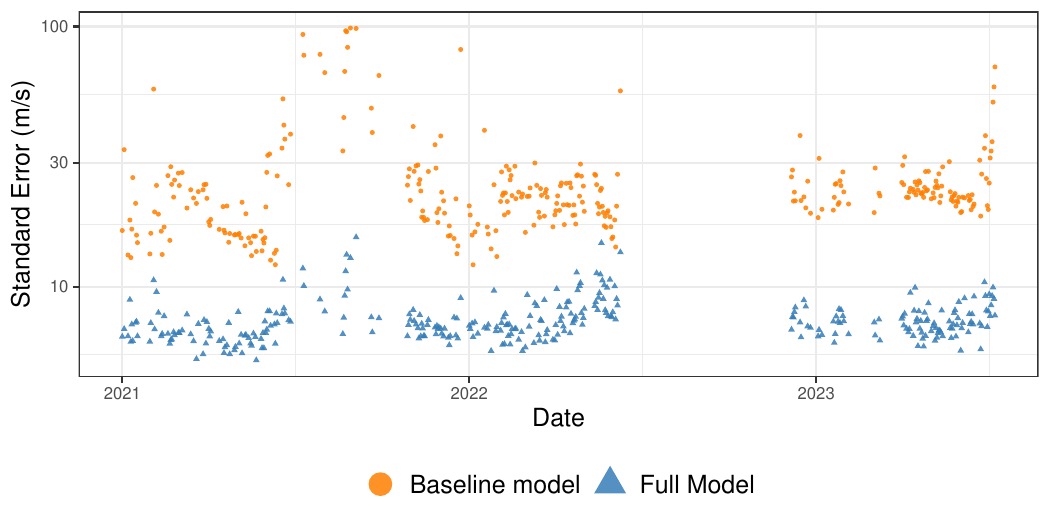}
    \caption{Time series of the line-by-line cleaned RV standard errors (see Equation \eqref{eq:standerror_cv} from the LOOMCV procedure on a log10 scale). The Baseline Model is in orange and the Full Model is in Blue.}
    \label{fig:cv_se}
\end{figure}

The standard errors of the line-by-line cleaned RVs, as described in Equation \eqref{eq:standerror_cv}, offer additional insights into the model performance.
These represent the variability in the line-by-line cleaned RVs and the cleaned RV estimates for each day.
Figure \ref{fig:cv_se} displays the standard error for the Baseline and Full Models resulting from the LOOMCV procedure.
On average, the Full Model reduces the standard error by 232\% compared to the Baseline Model for LOOMCV, 242\% for LOOWCV, and 248\% for LOODCV, underscoring its superior precision.
Furthermore, the standard error results for the Full Model and the Full Model with LASSO are nearly identical.

\subsection{Uncertainty Quantification Results}
\label{subsec:boot_results}

We estimate the standard error of the cleaned RV estimators, $\hat{\text{v}}_t^{\text{clean}}$, as described in \S\ref{subsec:mle}.
These estimates reflect the uncertainty in $\hat{\text{v}}_t^{\text{clean}}$ but rely on the strong assumption that the covariance matrix $\bm{\Sigma}$ is correctly specified.
Consequently, the standard errors derived using the MLE method are nearly identical across all days.
For the Full Model, these MLE standard errors range from 0.2396 to 0.2492, with a mean value of 0.2418.

\begin{figure}[htb!]
    \centering
    \includegraphics{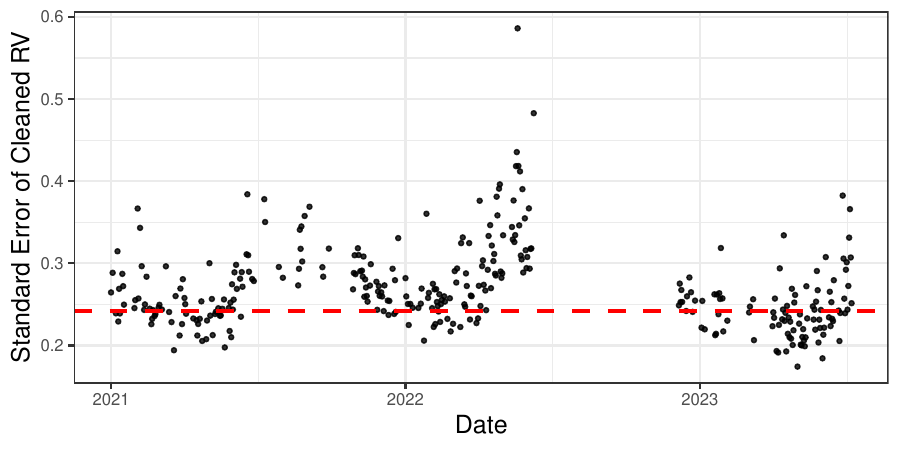}
    \caption{Standard errors for the cleaned RV from the Full Model. The black points represent bootstrap standard errors, while the red dashed black line is at the mean MLE standard error.}
    \label{fig:se_cleanRV}
\end{figure}

The wild bootstrap, described in \S\ref{subsubsec:bootstrap}, provides a potentially more accurate estimate of the uncertainty associated with $\hat{\text{v}}_t^{\text{clean}}$.
Due to its computational demands, we applied this approach only to the Full Model.
Using $R =$ 2,500 bootstrap samples, we use blocking to group the observations by line in order to capture autocorrelation.
The resulting standard errors are shown in Figure \ref{fig:se_cleanRV}.
The bootstrap standard errors are more variable across time than the MLE method; interestingly, they reflect similar trends as the CV standard errors (see Figure \ref{fig:cv_se}).

Finally, we used bootstrap samples of the model parameters to derive the sampling distribution of the RMSE, enabling an assessment of how well the planetary RV is recovered under heteroscedastic and autocorrelated errors.
For the Full Model, the 99\% bootstrap confidence interval for the RMSE is $(0.554,0.717)$.
This confidence interval is consistent with the RMSE estimates from the CV procedures (see Table \ref{tab:rmse_cv}).

\section{Discussion}
\label{sec:discuss}

We investigate the potential of a fixed effects linear model to mitigate the impact of stellar activity on contaminated RV estimates.
By focusing on 778 individual spectral lines, we estimate properties of their shapes, along with contaminated line-by-line RVs.
Our analysis demonstrates that the inclusion of line shape properties (e.g., depth, width, HG coefficients) in the proposed linear model significantly reduces RV variability due to sources such as stellar activity or instrumental effects, compared to models that do not include these covariates.
The significant reductions in AIC and BIC between the Baseline and Full Models supports this claim.
Using a fixed-effects framework, we produce estimates of both the cleaned RVs and their standard errors; these provide astronomers with valuable inputs for existing exoplanet detection methods (e.g., periodograms, fitting Keplerian models, Gaussian Processes).

Using day-averaged NEID solar measurements, we validate the proposed model's ability to recover the sun's COM RV signals.
This is contrasted with the model's predictive performance for the contaminated RVs, evaluated using metrics such as AIC, BIC, and RSE.
These metrics are particularly relevant for surveys targeting stars with unknown COM motion.
Our findings reveal that models with stronger predictive performance (for the line-by-line contaminated RVs) also trend to recover cleaned RVs more effectively (as determined via the RMSE metric).
There are some exceptions to this: (1) the Common Slopes Model has smaller AIC, BIC, and RSE but larger RMSE compared to the Baseline Model and (2) the Full Model w/ LASSO has smaller AIC and BIC but larger RMSE compared to the Full Model.
Overall, our findings suggest that developing even more predictive models could uncover small planetary signatures; though this may not hold universally.
We also note that this analysis uses solar data, and the relationship may not hold for observations from other stars.

The proposed model incorporates covariates that characterize the shape of the spectral lines; these include parameters derived from an inverted Gaussian-density shape and HG basis functions.
These provide measurements of every lines' overall shape and are used to control for sources of noise in the contaminated RV.
While using either parametrization improves predictive performance individually, combining them yields a significantly better model in terms of all examined performance metrics (AIC, BIC, RSE, RMSE).
Our model framework can easily incorporate additional line-level information, potentially further enhancing performance.

The large size and sparsity of our design matrices present both challenges and opportunities.
By taking advantage of modern libraries optimized for sparse matrix operations--such as quick multiplication and inversion--our models are fast to run. 
This computational efficiency makes it feasible to apply resampling and cross-validation techniques, which provides deeper insights into our models.

Our analysis also highlights the differential impact of stellar variability on individual spectral lines.
The model that constrained all lines to share the same slope for each covariate failed to clean the RV measurements as effectively as those allowing line-specific slopes.
It may be the case that subsets of lines may be impacted similarly by different sources of variability.
This possible clustering of lines may be accounted for in a statistical model; investigating this direction is left to future work.
Moreover, if there are stellar activity indicators that practitioners believe impact every line in the same way, our model framework can easily accommodate this scenario.

We also evaluate a regularized version of our model using LASSO. While this model slightly reduces the number of parameters, AIC, and BIC, it exhibits a minor increase in RMSE.
For most use cases, we recommend either the Full Model or the LASSO variant.
We particularly endorse the LASSO model if a large number of covariates are used, as it is more robust against over-fitting than traditional OLS estimators.

Our method provides a straightforward approach to control for differential RV offsets across distinct temporal groups.
For instance, our data span two time groups, before and and after a 6-month shutdown of NEID due to the fire.
We observed a negative bias in the estimated contaminated RVs of certain lines during the later time group compared to the earlier time group, but successfully controlled for this group-specific bias by including a time-group offset in the model.
A preliminary analysis suggested that allowing covariate slopes to vary across these groups marginally improved performance but significantly increased the number of model parameters.

Our analysis has a few limitations.
Specifying and estimating the covariance matrix remains challenging.
While OLS estimates under a general covariance structure converge to the true model parameters, they are not the most statistically efficient linear estimators; further complicating matters, the traditional OLS estimators for the standard errors of the model parameters become inconsistent when there is autocorrelation or heteroscedasticity.
Bootstrap methods partially address autocorrelation and heteroscedasticity, but incorporating the covariance structure explicitly into the model may yield superior results.

Another limitation is our treatment of time as a discrete variable, which results in $T$ parameters for the cleaned RV; this is contrasted with continuous-time models such as Gaussian Processes Regression or B-splines.
Additionally, the current model assumes linear relationships between covariates and RV, but non-linear effects of shape changes on RV may exist.
Exploring non-linear models and a continuous effect of time may further improve accuracy.

Overall, our models significantly reduce the effects of stellar activity on RV measurements, achieving a RMSE of 0.575 m s$^{-1}$.
Testing with cross-validation shows consistent performance, with RMSE ranging from 0.580 to 0.612 m s$^{-1}$ across different procedures.
The wild bootstrap analysis, which accounts for heteroscedasticity and autocorrelation, yields a 99\% confidence interval for RMSE of (0.554, 0.717).

These findings highlight the value of line-by-line analyses in reducing stellar activity contamination.
Further investigation into individual spectral lines and activity indicators may greatly enhance RV estimations.
We have made both our data along with the code publicly available.
The code is accessible at \url{https://github.com/Joesalzer/ExoRV_LM}, and the data can be found at \url{https://doi.org/10.5281/zenodo.14841435}.
\\
\\
\\
This material is based upon work supported by the U.S. National Science Foundation under Grant No. 2204701. 
E.B.F. contributions were partially supported by NASA Extreme Precision RV award \#80NSSC21K1035.
This work contains data taken with the NEID instrument, which was funded by the NASA-NSF Exoplanet Observational Research (NN-EXPLORE) partnership and built by Pennsylvania State University. NEID is installed on the WIYN telescope, which is operated by the National Optical Astronomy Observatory, and the NEID archive is operated by the NASA Exoplanet Science Institute at the California Institute of Technology. NN-EXPLORE is managed by the Jet Propulsion Laboratory, California Institute of Technology under contract with the National Aeronautics and Space Administration.

The Center for Exoplanets and Habitable Worlds is supported by Penn State and its Eberly College of Science.
The authors acknowledge the Penn State Institute for Computational and Data Science for providing computational resources and support that have contributed to the research results reported in this publication.

The Flatiron Institute is a division of the Simons Foundation. Support for this work was provided by NASA through the NASA Hubble Fellowship grant HST-HF2-51569 awarded by the Space Telescope Science Institute, which is operated by the Association of Universities for Research in Astronomy, Incorporated, under NASA contract NAS5-26555.

This work is based on observations at Kitt Peak National Observatory, NSF’s NOIRLab, managed by the Association of Universities for Research in Astronomy (AURA) under a cooperative agreement with the National Science Foundation. The authors are honored to be permitted to conduct astronomical research on Iolkam Du\'ag (Kitt Peak), a mountain with particular significance to the Tohono O\'odham.

\facilities{WIYN}

\software{R \citep{r_cite}, tidyverse \citep{tidyverse}, HDF5 \citep{rhdf5}, Matrix \citep{matrix}, patchwork \citep{patchwork}, collapse \citep{rcollapse}, pbmcapply \citep{pbmcapply}.}

\bibliography{exorv}

\end{document}